\documentclass[preprint,5p]{elsarticle}
\usepackage{braket}
\usepackage{cancel}
\usepackage{titlesec}
\usepackage[utf8]{inputenc}
\usepackage{dcolumn}
\usepackage{multirow}
\usepackage{here}
\usepackage{color}
\usepackage{amsmath}
\usepackage{lineno}
\usepackage{geometry}

\begin{document}
\biboptions{sort&compress}
\title{Attosecond-resolved coherent control of zone-folded acoustic phonons in silicon carbide}
\journal{\empty}

\author[{1}]{Hiromu Matsumoto}
\author[{1}]{Tsukasa Maruhashi}
\author[{1,2}]{Yosuke Kayanuma}
\author[{3,4}]{Yadong Han}
\author[{3,4}]{Jianbo Hu}
\author[{1,5,6}]{Kazutaka G. Nakamura}
\ead{nakamura@msl.titech.ac.jp}

\affiliation[1]{organization={Laboratory for Materials and Structures, Institute of Innovative Research, Tokyo Institute of Technology},
addressline={4259 Nagatsuta},
city={Yokohama},
postcode={226-8501},
country={Japan}}

\affiliation[2]{organization={Graduate School of Science, Osaka Metropolitan University},
addressline={1-1 Gauken-cho, Sakai},
city={Osaka},
postcode={599-8531},
country={Japan}}

\affiliation[3]{organization={State Key Laboratory for Environment-Friendly Energy Materials, Southwest University of Science and Technology},
addressline={Mianyang},
city={Sichuan},
postcode={621010},
country={China}}

\affiliation[4]{organization={Laboratory of Shock Wave and Detonation Physics Research, Institute of Fluid Physics, China Academy of Engineering Physics},
addressline={Mianyang},
city={Sichuan},
postcode={621900},
country={China}}

\affiliation[5]{organization={Materials and Structures Laboratory, Institute of Integrated Research, Institute of Science Tokyo},
addressline={4259 Nagatsuta},
city={Yokohama},
postcode={226-8501},
country={Japan}}

\affiliation[6]{organization={SIT Research Laboratories, Shibaura Institute of Technology},
addressline={3-7-5 Toyosu, Koto-ku},
city={Tokyo},
postcode={135-8548},
country={Japan}}

\cortext[cor1]{Corresponding author}

\begin{abstract}
Zone-folded acoustic phonons \textcolor{black}{(6 THz)} in 4H silicon carbide (SiC) have been coherently excited using a femtosecond near-infrared pulse and measured through transient reflectivity with a pump and probe protocol.  
Their amplitude is coherently controlled with \textcolor{black}{300-attoseconds} precision and the results \textcolor{black}{show interference fringe patterns due to electronic and phonon interference.
The results are} well reproduced by a model calculation with two electronic and phonon levels and an impulsive stimulated Raman process.
Using the model, we obtain the analytical form of the coherent control scheme at an off-resonant condition.
\end{abstract}


\maketitle


\section{Introduction}

Coherent control is a technique to manipulate quantum states using optical pulses \cite{Shapiro2003} and is widely used to control  electronic states, vibrational states and spins in condensed matter. Optical phonons are excited and detected by using femtosecond optical pulses and its amplitude is controlled using a pair of the pulses \cite{Weiner1993, Dekorsy1993, Hase1996}.
Its delay has been usually controlled within the vibrational period (several tens of femtoseconds) to control phonons.
Furthermore, recently the generation of phonons including electronic transition processes can be controlled using pulses controlled with attosecond precision \cite{Nakamura20192, Kimata2020, Yu2022, Takagi2023}.

Ultrafast dynamics of coherent phonons in \textcolor{black}{silicon carbide (SiC) has been} studied using a femtosecond pulses \cite{Kato2012, Arashida2018}, \textcolor{black}{because SiC has a wide band gap and high thermal conductivity and is expected to be applied to power devices \cite{She2017}.} 
Phonons are coherently excited by ultrashort optical pulses, the pulse width of which is shorter than the vibrational periods, and detected as a transient change in transmissivity or reflectivity \cite{Nelson1985, Zeiger1992, Merlin1997,Dekorsy2000, Nakamura2015}. These coherently excited phonons are called coherent phonons and used to study their dynamics.
The coherent phonons of the longitudinal optical (LO) mode with a frequency of approximately 29 THz have been excited and detected as a transient change in reflectivity\cite{Kato2012} using a sub-10-fs pulse. The coherent phonons of the folded transverse optical (FTO) mode at 23.3 THz and the folded transverse acoustic (FTA) mode at 6.1 THz have been excited and detected in transient transmission experiments with sub-10-fs optical pulses. Furthermore, their amplitudes were controlled using a pulse-shaping technique \textcolor{black}{with a spatial light modulator} \cite{Arashida2018}.  However, the coherent FTA-mode oscillation was barely discernable in transient signals, being immersed by the intense optical phonon oscillation.

\color{black}
The low-frequency FTA phonons can be excited by a 50-fs infrared pulse, which is a well-defined Gaussian pulse. 
It oscillation can be easily resolved from those of electronic states.
Therefore, the FTA phonons \textcolor{black}{are} suitable  for studying a coherent phonon control scheme in off-resonant conditions.
\color{black}

\color{black}
In this paper, we report our study of coherent control of FTA-mode phonons in 4H-SiC using a pair of relative-phase-locked femtosecond pulses with precision of 300 attoseconds.
The results are analyzed using a theoretical model  involving two electronic levels and phononic levels, an impulsive stimulated Raman process, and the density matrix formalism.
\color{black}

\section{Experimental}

The coherent-phonon oscillation was detected as a change in reflectance of the femtosecond pulse using a pump-probe protocol \cite{Nakamura20192}.
The laser used was a Ti:sapphire oscillator (MTS-a, KMLabs), which generated femtosecond pulses (center wavelength = 798 nm (1.54 eV), spectral width of 28 nm (full-width at half-maximum) and a pulse width of approximately 42.5 fs). The pulse from the oscillator passed through a pair of chirp mirrors to compensate the group-velocity dispersion of the optics. 
\textcolor{black}{The pulse was split with a partial beam splitter into two pulses and used as pump and probe pulses.}
The pump pulse was introduced into a scan delay unit to control the delay between the pump and probe pulses. The pump and probe power were 25 and 7.8 mW, respectively. 
The pump pulse was then introduced into a \textcolor{black}{custom built} Michelson-type interferometer and split into a pulse pair, the relative phases of which were locked (pump 1, pump 2). One optical arm of the interferometer was equipped with an automatic positioning stage, which has an active-feedback controlled by a controller with a minimum resolution of 5 nm and a repetition position accuracy of $\pm$10 nm.

\textcolor{black}{In the double-pulse experiments, we controlled the stage with in steps of 45 nm, which corresponds to an optical-path change of 90 nm and the delay approximately 300 attoseconds.}
The estimated phase stability was approximately 0.049 $\pi$ for 800-nm light.  
The interval between pumps 1 and 2 was characterized with an optical interference and frequency-resolved autocorrelation (FRAC) measurements.
The relative phase-locked pump pulses (pump 1 and 2) and the delayed probe pulse (hereafter refereed to as pulse 3) were focused on a sample using a lens. The pump pulses were linearly polarized and their polarizations were mutually orthogonal.  
\textcolor{black}{The reflected probe pulse was detected using electro-optical (EO) sampling. 
The signal was obtained by accumulating the measured values 4000 times.} 
The sample used was a single crystal of hexagonal 4H-SiC (0001) obtained from SICC Co. Ltd. Its resistivity is more than $1 \times 10^7 \  \Omega \cdot$cm. The measured thickness of the sample was $420 \pm 10 \  \mu$m.

 
\section{Results and discussion}

\subsection{Single-pulse-excitation experiment}
Figure 1 shows the transient reflectivity of the probe pulse after irradiating only pump 1 (single-pulse experiment). 
\textcolor{black}{The data were obtained by averaging 40 measurements.}
After the intense response at delay zero, there is a very weak oscillation.  
The vertically enlarged spectrum (Fig. 1(b)) clearly shows a coherent oscillation.

\begin{figure}[h]
\includegraphics[width=8.5 cm]{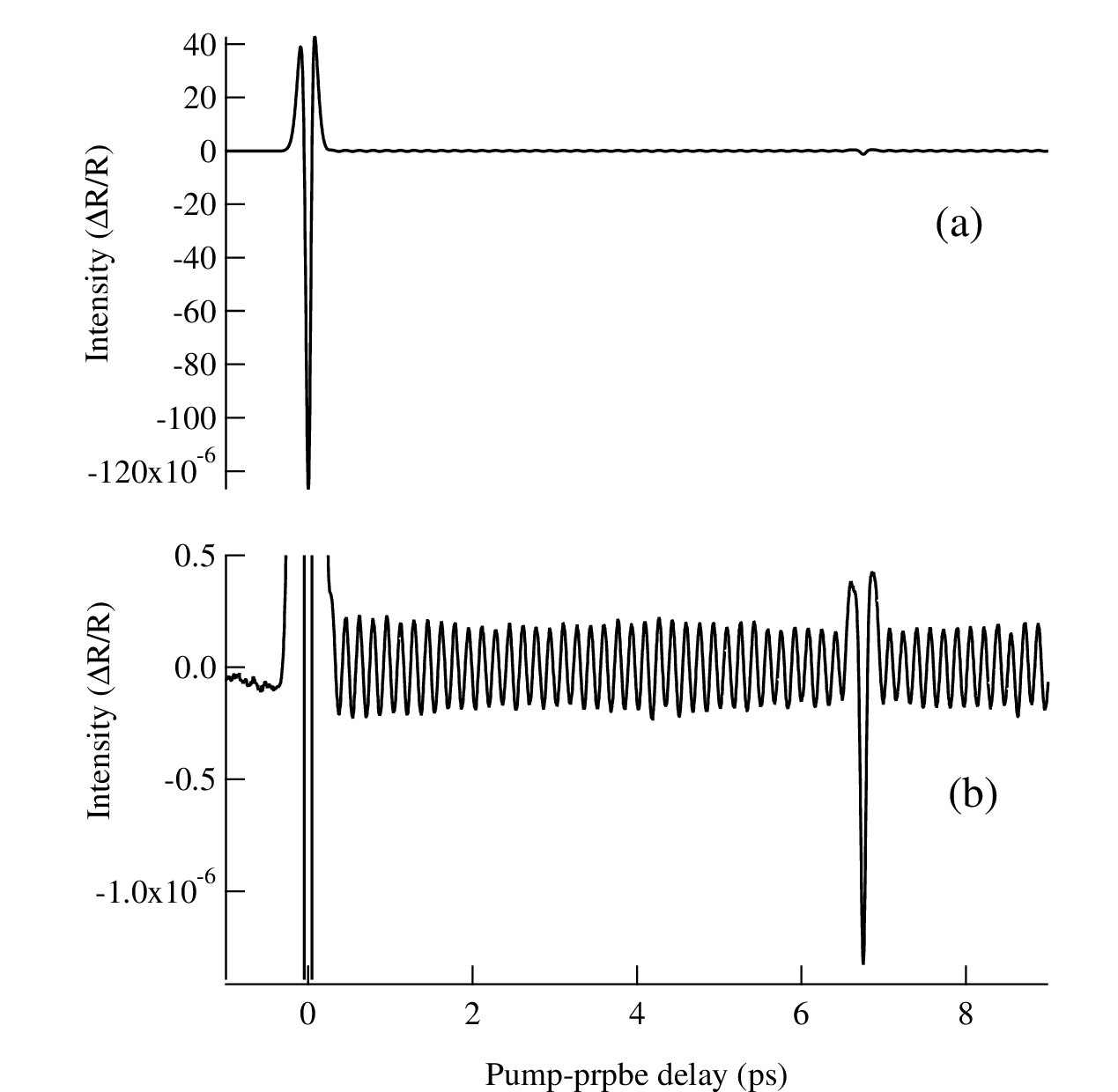}
\caption{(a) Intensity of the reflected probe pulse along a pump probe delay, and (b) its vertically enlarged spectrum.} 
\label{SiC1}
\end{figure}

The coherent oscillation after the strong peak at zero delay was analyzed with a sinusoidal function $A \sin (\omega t + \theta)$ for small regions with a width of approximately 300 fs in the range between 0.4 and 5.8 ps. Figure 2(a) and (b) shows respectively the oscillation amplitudes and frequencies obtained for every 300 fs.
The oscillation frequency is almost constant and estimated to be $6.0 \pm 0.2$ THz.
The oscillation was assigned to a coherently excited FTA mode with $q/q_B =2/4$ in 4H-SiC by comparing the Raman spectrum (see \textcolor{black}{Supplemental A}). In the Raman spectrum, the FTA mode was observed at 203 cm$^{-1}$, which corresponds to 6.09 THz.

\begin{figure}[h]
\begin{center}
\includegraphics[width=7.5 cm]{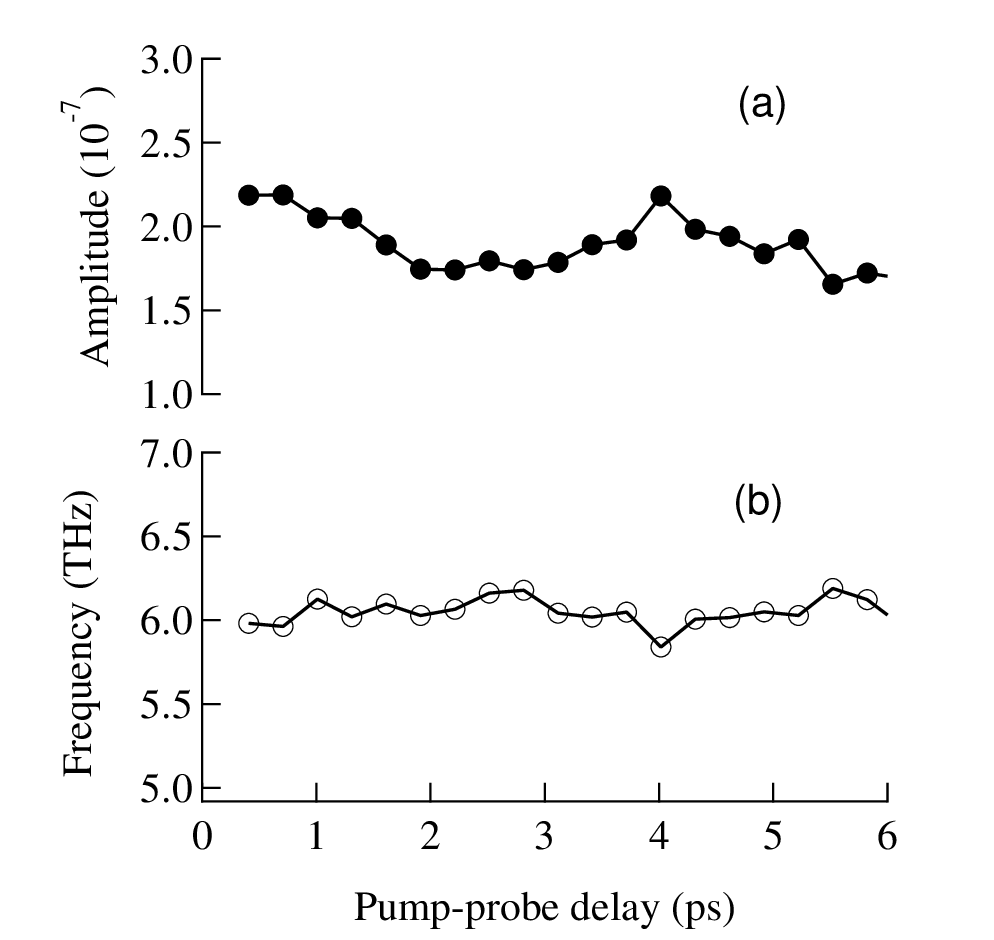}
\end{center}
\caption{(a) The amplitude, and (b) the frequency obtained by curve fitting the transient reflectivity (shown in Fig. 1) with a sinusoidal function.} 
\label{amp1}
\end{figure}

\textcolor{black}{The oscillation amplitude does not show remarkable decay within 6 ps (Fig. \ref{amp1} (a)). Then the lifetime of the FTA mode is much longer than 6 ps.}

The strong peak at delay zero is a coherent artifact because of overlapping pump and probe pulses.
The spike at approximately 6.75 ps may be a coherent artifact because of an overlapping reflected pump pulse from the rear side of the sample with the probe pulse. The intensity of this peak is approximately 1/100 compared to the peak at time zero.
The sample thickness is $420 \pm 10 \ \mu$m, and the refractive index of a silicon carbide crystal is approximately 2.6 for light of wavelength 800-nm.\cite{Sibgh1971}   Therefore, the propagation time is \textcolor{black}{estimated to be} about 7.28 ps.

\subsection{Coherent control experiment with double-pulse excitation}

The FTA-mode phonon was coherently controlled by the double-pulse excitation with a pair of relative phase-locked femtosecond pulses.
Figure 3 shows a two-dimensional map of the transient reflectivity plotted against the pump-pump delay ($t_{12}$; vertical axis) and pump1-probe delay ($t_{13}$; horizontal axis).
The FTA-mode-phonon oscillations are observed along the horizontal axis. Their amplitudes are suppressed at a pump-pump delay of around 80 and 240 fs.
The coherent phonon amplitude of the FTA mode was obtained from the peak intensity at approximately 6.0 THz in the Fourier-transformed spectrum of the transient reflectivity along the pump-probe delay in a range between 0.72 and 2.06 ps after the second pump pulse for each pump-pump delay.  

\begin{figure}[h]
\begin{center}
\includegraphics[width=8cm]{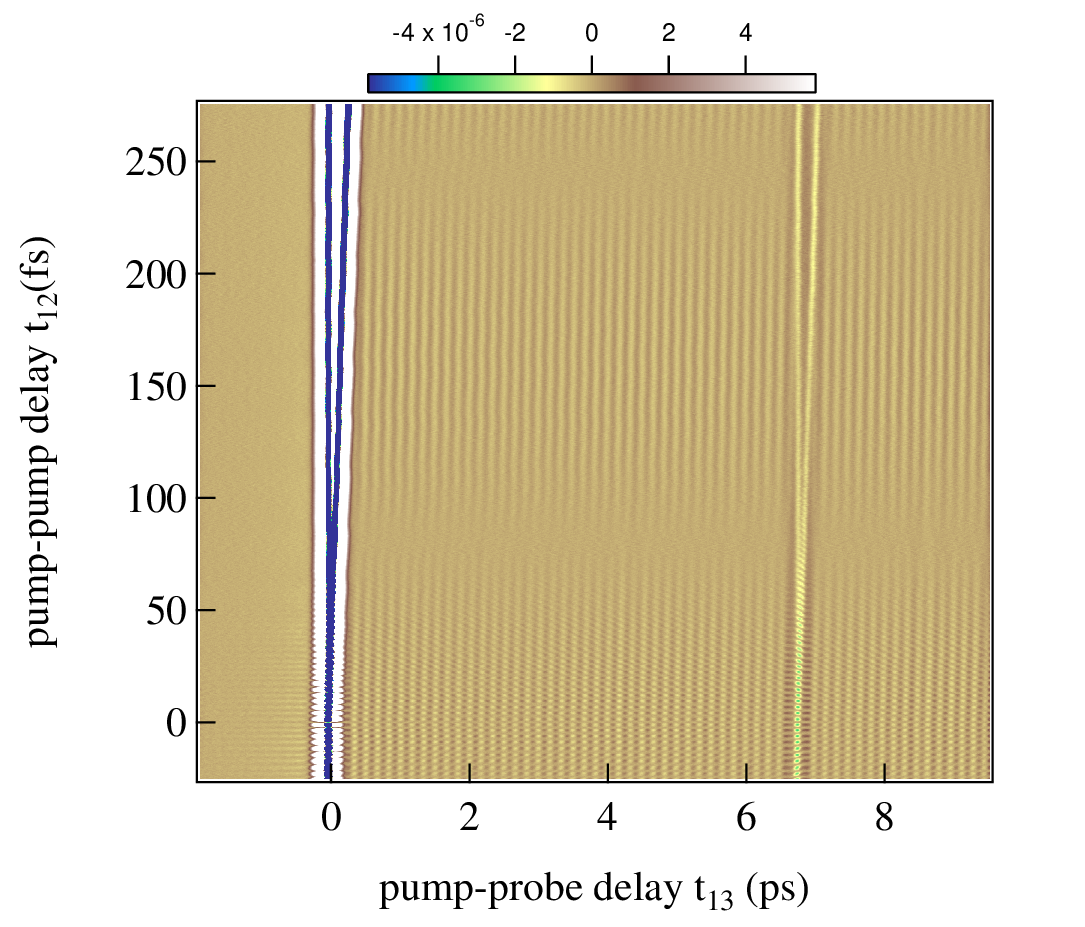}
\end{center}
\caption{Two-dimensional map of the transient reflectivity plotted against the pump-pump delay ($t_{12}$; vertical axis) and pump1-probe delay ($t_{13}$; horizontal axis).} 
\label{SiC2}
\end{figure}

The phonon amplitude of the FTA mode along the controlled pump-pump delay [Fig. 4(a)] shows enhancements at $t_{12} \approx 0$ and $t_{12} \approx 160$ and suppressions at $t_{12} \approx 80$ and $t_{12} \approx 240$. The maximum amplitude at $t_{12} \approx 0$ is two times larger than that at $t_{12} \approx 160$.
These features of enhancement and suppression of the phonon amplitude may arise from interference of the coherent phonon oscillations of the FTA mode.

In addition to the phonon interference pattern, there is another oscillation with a period of approximately 2.7 fs and upwards to approximately 80 fs. The oscillation period has the same value as the optical cycle of the \textcolor{black}{near-infrared} pulse. 
\textcolor{black}{The optical interference pattern and the FRAC of the two pump pulses are shown in Fig. 4 (b) and (c), respectively.
The optical interference is superimposed on the interference the coherent TFA-mode phonons due to the mediation of electronic states within the material.}

\begin{figure}[h]
\includegraphics[width=8.5 cm]{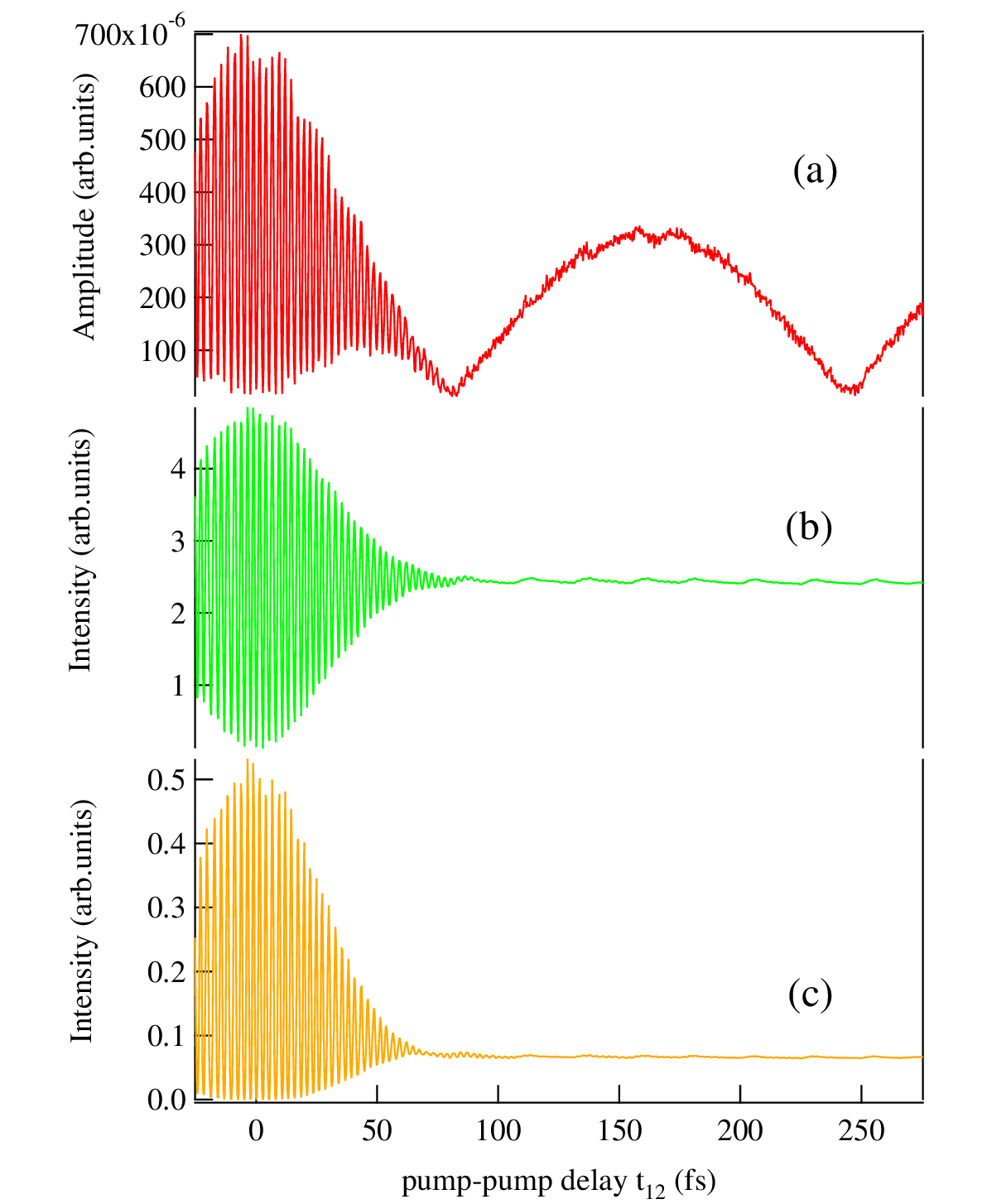}
\caption{(a) Amplitude of the controlled oscillation after pump 2, (b) the optical interference, and (c) the FRAC against the pump-pump delay ($\tau$). The amplitude is normalized using that obtained after excitation after pump 1 only; oscillation between the irradiation timing of pumps 1 and 2.} 
\label{SiC3}
\end{figure}

\subsection{Theoretical calculation for coherent control}

\textcolor{black}{Coherent optical phonons are excited by an ultrashort optical pulse, the width of which is much shorter than the oscillation period of the phonon \cite{Nelson1985, Dekorsy2000}.
The generation mechanism has been explained in terms of the displacive excitation of coherent phonons for absorbing materials \cite{Zeiger1992} and impulsive stimulated Raman scattering (ISRS) for transparent materials \cite{Nelson1985}.
Both generation processes can be handled uniformly using a quantum mechanical model with two electronic levels coupled to harmonic oscillators and the density matrix formalism \cite{Nakamura2015}.
}
In the present experimental conditions, the energy of the pump pulse ($\sim 1.55$ eV) is well below the band gap ($\sim 3.26$ eV) of the 4H-SiC sample. Then, ISRS is the main generation process for coherent phonons in 4H-SiC. For the ISRS process and a Gaussian pulse, we can simplify the model calculation and obtain an analytical form for coherent control of the phonons.

\color{black}
We model a simple system consisting of two electronic states and a harmonic oscillator for phonon \cite{Nakamura2015, Kimata2020, Nakamura2024}.
The electronic ground and excited states are denoted by $\ket{g}$ and $\ket{e}$, respectively.
The Hamiltonian of the system is given by
 \begin{align}
 \hat{H}_0 = \hbar \epsilon \ket{e}\bra{e} + \hbar \omega \hat{b}^\dagger \hat{b} + \alpha \hbar \omega (\hat{b}^\dagger + \hat{b}) \ket{e}\bra{e},
\end{align}
where $\hbar \epsilon$ is the energy between the electronic ground and excited states (band gap energy), $\omega$ is the angular frequency of the harmonic oscillator (phonon), $\alpha$ is the electron-phonon coupling constant. 
$\hat{b}$ and $\hat{b}^\dagger$ are annihilation and creation operators of the phonon, respectively. 
The interaction Hamiltonian of the electronic states with light is obtained with the dipole interaction and the rotating wave approximation:
 \begin{align}
 \hat{H}_I (t) &= \mu \sum_{i=1}^2 ( E_i(t) \ket{e}\bra{g} + E_i^*(t) \ket{g}\bra{e}), 
\end{align}
where $\mu$ is the electronic transition dipole moment, $E_i(t)$ is an electric field of the $i$-th pulse  for a pair of phase-locked pulses ($i=1,2$).
Here we assume the Gaussian pulse:
  \begin{align}
 E_i(t) &= \frac{E_0}{\sqrt{\pi} \sigma} \exp(- \frac{(t-\tau_i)^2}{\sigma^2}) e^{i \Omega (t-\tau_i)},
 \end{align}
 where $\Omega$ is the center angular frequency of the pulse, $\sigma$ is the pulse width and $\tau_i$ is a delay $\tau_1 =0$ and $\tau_2 = \tau$ between two pulses.

The initial state of the system is set to $\ket{g, 0} = \ket{g} \otimes \ket{0}$, where $\ket{0}$ is the phonon vacuum state.
By applying the perturbation theory, the density operator at time $t$ after the second pulse excitation is obtained by
  \begin{align}
 \hat{\rho}_{jk} (t) &=  \alpha   \Big(\frac{\mu}{ \hbar} \Big)^2 e^{-i \omega t} e^{i \Omega (\tau_j - \tau_k)} \nonumber\\
&  \int_{-\infty}^{t_2}  \int_{t_1}^{t} \exp \Big(-\frac{(t_1-\tau_j)^2+(t_2-\tau_k)^2}{\sigma^2} \Big) \nonumber\\
& e^{-i(\epsilon-\Omega)(t_2-t_1)} (e^{i\omega t_2} - e^{i \omega t_1}) dt_2 dt_1 \ket{g,1}\bra{g,0}
 \end{align}

 In off-resonance, the time the system spends in an electronic coherence is limited by the Heisenberg relation ($\Delta t \approx 1/\Delta \omega$, with a detuning $\Delta \omega = (\epsilon - \Omega)$) \cite{Mukamel1995}.
Then the coherent time of the polarization is short and $t_1$ is approximated by $t_1 = t_2 -\delta$ with a small value $\delta$.
Then we get
\begin{align}
e^{i\omega t_2} - e^{i\omega t_1} = e^{i\omega t_2} (1-e^{-i \omega \delta}) \approx (-i \omega \delta) e^{i \omega t_2}.
\end{align}
 
 Then the density operator is obtained as
 \begin{align}
\hat{\rho}_{jk}(t)  &  \approx  (-i \omega \delta) A  \exp (i \Omega (\tau_j - \tau_k)) \nonumber\\
\times &  \exp \Big(- i \omega (t-\frac{\tau_j +\tau_k}{2} + \theta)  \Big) \nonumber\\
  \times& \exp \Big( -\frac{(\tau_k -\tau_j)^2}{2 \sigma^2} \Big)  \ket{g1}\bra{g0},
 \end{align}
 where the phase $\theta$ is defined as $\theta = (\epsilon -\Omega)\delta/\omega$, and
 the real value $A$ is defined as
  \begin{align}
  A = \alpha \Big( \frac{\mu}{\hbar} \Big)^2   \sqrt{\frac{\pi}{2}} \sigma \exp \Big(-\frac{\omega^2 \sigma^2}{8} \Big).
  \end{align}
 
Then we get
\color{black}
 \begin{align}
\sum_{j,k=1}^2 \hat{\rho}_{jk} (t) =&  -i \omega \delta \exp \Big(-i \omega (t- \frac{\tau}{2}+ \theta) \Big) \nonumber\\
&\times  \Big(2 \cos (\frac{\omega \tau}{2}) + B \Big) \ket{g1}\bra{g0},
  \end{align}
\color{black}  
  where the real value $B$ is defined as
  \begin{align}
  B = 2 \cos (\Omega \tau) \exp (-\frac{\tau^2}{2 \sigma^2}).
  \end{align}
  
  The total density operator $\rho (t)$ is
 \begin{align}
\hat{\rho} (t) = \sum_{j,k=1}^2 \Big(  \hat{\rho}_{jk} (t) + \hat{\rho}_{jk}^\dagger (t) \Big).
  \end{align}
  
 The atomic displacement is expressed by an operator $\hat{Q} = C(\hat{b}^\dagger + \hat{b})$, where $C$ is the constant of proportionality.
The expectation value of the atomic displacement is obtained by 
 \begin{align}
\braket{Q(t)} =& {\rm{Tr}} [\hat{Q} \hat{\rho} (t)] \nonumber\\
=&  2C \omega \delta  \Big(2 \cos (\frac{\omega \tau}{2}) + B \Big) \sin \Big(\omega (t - \frac{\tau}{2}+ \theta) \Big).
 \end{align}
 Then it oscillates with an angular frequency of $\omega$ and a phase of $(-\omega \tau/2+\theta)$: $\braket{Q(t)} = Q_0 \sin (\omega t -\omega \tau/2+\theta)$.
 The phonon amplitude $Q_0$ is obtained by
  \begin{align}
Q_0 =&  \Big| 2C \omega \delta  \Big(2 \cos (\frac{\omega \tau}{2}) + B \Big) \Big| \nonumber\\
=& 2C\omega \delta \sqrt{2(1+\cos (\omega \tau)) + 4B \cos (\omega \tau/2) + B^2}.
 \end{align}
\textcolor{black}{ (More detailed calculations are shown in Supplemental B)}


\color{black}

The first two terms represent the phonon interference between phonons induced by the first and second pumps. The third and fourth terms include the optical interference between the pumps.  
\begin{figure}[htbp]
\includegraphics[width=7.5 cm]{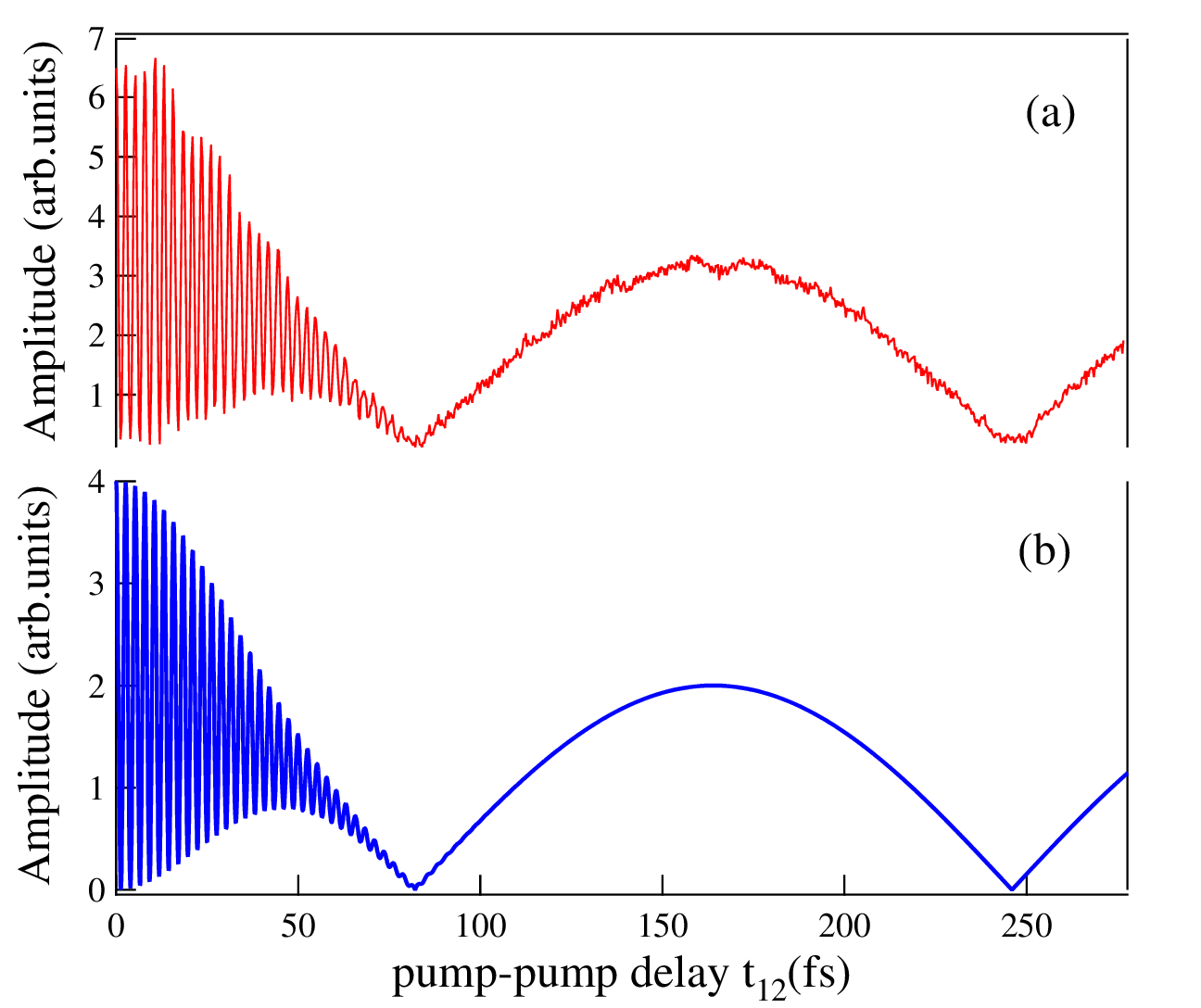}
\caption{Phonon amplitude after pump 2 irradiation plotted against the pump-pump delay. (red curve (a)) experimental results, which is the same as in Fig. 4(a), and (blue curve (b)) simulation using equation (7) given in the text.} 
\label{SiC4}
\end{figure}

The calculated phonon amplitude is shown in Fig. \ref{SiC4} as a function of the pump-pump delay for $\omega = 2\pi/164$ for the FTA-mode phonon of 4H-SiC, $\Omega = 2\pi/2.63$, and $\sigma = 27$ fs.
The simulation reproduces well the experimental result.
The fast oscillation with period of approximately 2.7 fs is the optical interference superimposed on the phonon interference.

In summary, the coherent phonons with the FTA mode (at 6.0 THz) were excited and clearly observed using 42-fs near-infrared pulses.  
The amplitude of the FTA-mode phonons was coherently controlled \textcolor{black}{with attosecond precision} using relative phase-locked pulses. We derived the analytical form of the coherent control assuming a short electronic coherence time. The experimental results of the amplitude control were well explained by the derived analytical form including electronic interference.

\section*{Acknowledgments}
The authors thank Tetsuya Kimata of Tokyo Institute of Technology for measuring Raman spectra.
\textcolor{black}{The authors thank Ikufumi Katayama of Yokohama National University and Jun Takada of Shibaura Institute of Technology for valuable discussion.}
This work was supported in part by JSPS KAKENHI Grant Number JP15K13377, JP17K19051, JP17H02797, JP19K03696, JP19K22141, \\
JP21K18904, JP21H0449, JP22H01984, JP23K23252, and JP25H00829, Collaborative Research Project of Laboratory for Materials and Structures, Institute of Innovative Research, Tokyo Institute of Technology.




\vspace{0.5cm}
\noindent {\bf{CRediT authorship contribution statement}}\\

{\bf{Hiromu Matsumoto:}} Writing - review, Writing - original draft, Investigation, Formal analysis, Data curation.
{\bf{Tsukasa Maruhashi:}} Data curation.
{\bf{Yosuke Kayanuma:}} Writing - review, Formal analysis.
{\bf{Yadon Han:}}  Investigation, Conceptualization.
{\bf{Jianbo Hu:}} Writing - review, Conceptualization.
{\bf{Kazutaka G. Nakamura:}} Writing - review \& editing, Writing - original draft, Formal analysis, Funding acquisition, Conceptualization.
\color{black}

\vspace{0.5cm}
\noindent {\bf{Declaration of generative AI and AI-assisted technologies in the manuscript preparation process}}\\

During the preparation of this work the authors used DeepL in order to improve language of the manuscript. After using this tool/service, the authors reviewed and edited the content as needed and take full responsibility for the content of the published article.
\color{black}



\pagebreak
\onecolumn

\newgeometry{top=25mm, bottom=25mm, left=25mm, right=25mm}

\begin{center}
{\bf{\large{Supplemental Materials: Attosecond-resolved coherent control of zone-folded acoustic phonons in silicon carbide}}}\\
\vspace{8mm}
Hiromu Matsumoto$^1$, Tsukasa Maruhashi$^1$, Yosuke Kayanuma$^{1,2}$, \\
Yadon Han$^{3,4}$, Jianbo Hu$^{3,4}$, and Kazutaka G. Nakamura$^{1,5,6}$\\
\vspace{8mm}
{\small{$^1$\textit{Laboratory for Materials and Structures, Institute of Innovative Research, Tokyo Institute of Technology, 4259 Nagatsuta, Yokohama, 226-8501, Japan}}}\\
{\small{$^2$\textit{Graduate School of Science, Osaka Metropolitan University, 1-1 Gauken-cho, Sakai, Osaka, 599-8631, Japan}}}\\
{\small{$^3$\textit{State Key Laboratory for Environment-Friendly Energy Materials, Southwest University of Science and Technology, Mianyang, Sichuan, 621010, China}}}\\
{\small{$^4$\textit{Laboratory of Shock Wave and Detonation Physics Research, Institute of Fluid Physics, China Academy of Engineering Physics, Mianyang, Sichuan, 621900, China}}}\\
{\small{$^5$\textit{Materials and Structures Laboratory, Institute of Integrated Research, Institute of Science Tokyo, 4259 Nagatsuta, Yokohama, 226-8501, Japan}}}\\
{\small{$^6$\textit{SIT Research Laboratories, Shibaura Institute of Technology, 3-7-5 Toyosu, Koto-ku, Tokyo, 135-8548, Japan}}}
\end{center}

\setcounter {equation}{0}
\setcounter {figure}{0}
\setcounter {table}{0}
\setcounter {page}{1}
\makeatletter
\renewcommand{\thefigure}{S\arabic{figure}} 
\renewcommand{\theequation}{S\arabic{equation}} 

\vspace{0.8cm}
\section*{A: Raman spectrum}

\begin{figure}[h]
\begin{center}
\includegraphics[width=4.5cm]{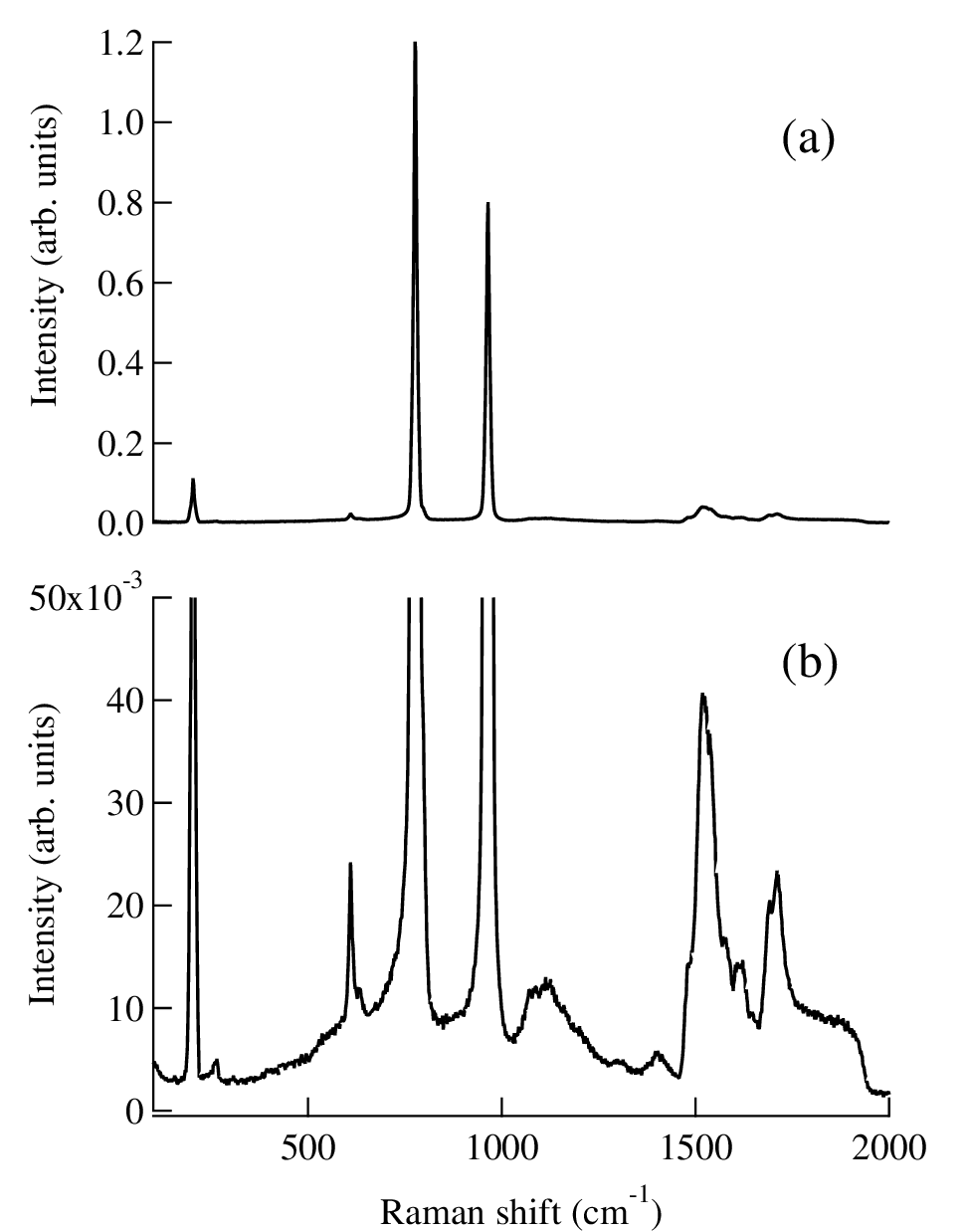}
\end{center}
\caption{(a) Raman spectrum of semi-insulating 4H-SiC taken with 532 nm excitation; (b) the spectrum is also shown with a rescaled coordinate to show detail. }
\label{A1}
\end{figure}

Figure \ref{A1} shows the Raman spectrum of the present sample (4H-SiC) obtained using a Raman spectrometer (LabRAM HR Evolution, HORIBA Ltd.) over the wavelength range of 500-2000 cm$^{-1}$ at a resolution of 1.65 cm$^{-1}$.  
Estimating the lifetime of a sharp peak is hard, because the spectral resolution is limited.
Incident radiation at 532 nm was provided using a single mode green laser (JUNO J050GS-16, Showa Optronics) at power below 5 mW.  The Raman spectrum was obtained with a back-scattering geometry.  
There are three strong peaks at 203, 776, and 964 cm$^{-1}$ (Fig. \ref{A1}(a)), weak peaks at 265 and 609 cm$^{-1}$ and weak and broad peaks at around 1113, 1520, and 1710 cm$^{-1}$ (Fig. \ref{A1}(b)). The observed spectrum is in good agreement with that in the literature.\cite{Nakashima1997, Burton1998, Burton1999, Tunhuma2019}
The major peaks at 203, 776, and 964 cm$^{-1}$ are assigned respectively to a planar FTA mode with $q/q_B=2/4$, a planar folded transversal optical (FTO) mode with  $q/q_B=2/4$, and an axial folded longitudinal optical (FLO) mode with $q/q_B=0$.\cite{Nakashima1997}   The 203 cm$^{-1}$ frequency corresponds to 6.09 THz.
The peaks at  265 and 609 cm$^{-1}$ can be assigned to a FTA mode with $q/q_B=4/4$ and a FLA mode with $q/q_B=4/4$, respectively.\cite{Nakashima1997}  The peaks at around  1520, and 1710 cm$^{-1}$ are second-order Raman spectral peaks and reported to be assigned to an overtone modes of the optical branch.\cite{Burton1999}

\color{black}

\section*{B: Quantum model with two electronic levels with harmonic oscillators}

We model a simple system consisting of two electronic states and a harmonic oscillator for phonon. \cite{Nakamura2015, Kimata2020, Nakamura2024}
The electronic ground and excited states are denoted by $\ket{g}$ and $\ket{e}$, respectively.
The Hamiltonian of the system is given by
 \begin{align}
 \hat{H}_0 = \hbar \epsilon \ket{e}\bra{e} + \hbar \omega \hat{b}^\dagger \hat{b} + \alpha \hbar \omega (\hat{b}^\dagger + \hat{b}) \ket{e}\bra{e},
\end{align}
where $\hbar \epsilon$ is the energy between the electronic ground and excited states (band gap energy), $\omega$ is the angular frequency of the harmonic oscillator (phonon), $\alpha$ is the electron-phonon coupling constant. 
$\hat{b}$ and $\hat{b}^\dagger$ are annihilation and creation operators of the phonon, respectively. 
The interaction Hamiltonian of the electronic states with light is obtained with the dipole interaction and the rotating wave approximation:
 \begin{align}
 \hat{H}_I (t) &= \mu \sum_{i=1}^2 ( E_i(t) \ket{e}\bra{g} + E_i^*(t) \ket{g}\bra{e}), 
\end{align}
where $\mu$ is the electronic transition dipole moment, $E_i(t)$ is an electric field of the $i-$th pulse  for a pair of phase-locked pulses ($i=1,2$).
Here we assume the Gaussian pulse:
  \begin{align}
 E_i(t) &= \frac{E_0}{\sqrt{\pi} \sigma} \exp(- \frac{(t-\tau_i)^2}{\sigma^2}) e^{i \Omega (t-\tau_i)},
 \end{align}
 where $\Omega$ is the center angular frequency of the pulse, $\sigma$ is the pulse width and $\tau_i$ is a delay $\tau_1 =0$ and $\tau_2 = \tau$ between two pulses.

\begin{figure}[h]
\vspace{-0.5cm}
\begin{center}
\includegraphics[width=2.5cm]{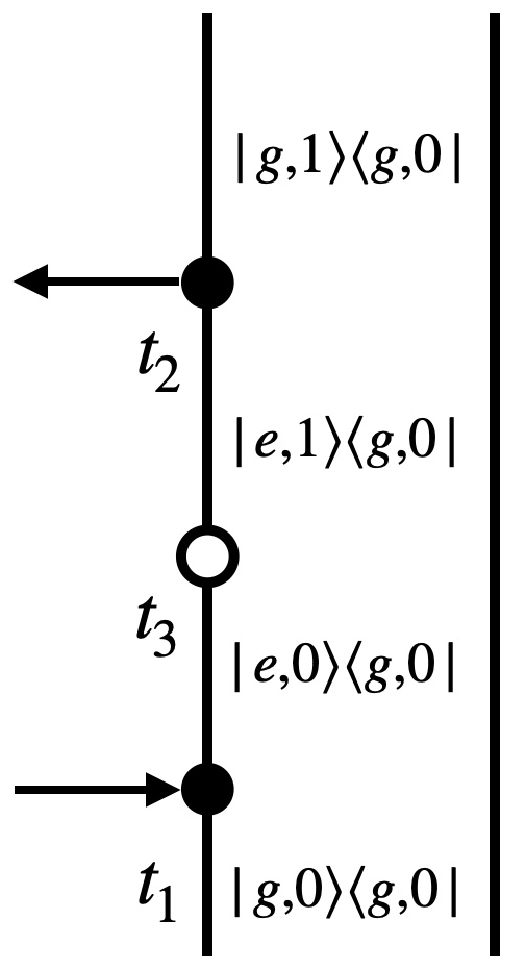}
\end{center}
\vspace{0.5cm}
\caption{Double-sided Feynman diagram. Black and white circles represent light-matter interaction and electron-phonon interaction, respectively.}
\label{A2}
\end{figure}

The initial state of the system is set to $\ket{g, 0} = \ket{g} \otimes \ket{0}$, where $\ket{0}$ is the phonon vacuum state.
By applying the perturbation theory, the density operator at time $t$ after the second pulse excitation (the excitation pulse $E_j(t_1)$ and the de-excitation pulse $E_k(t_1)$) is obtained by
  \begin{align}
 \hat{\rho}_{jk} (t) &=  \int_{-\infty}^{t_2}  \int_{t_1}^{t} \int_{t_1}^{t_2}  \Big(\frac{\mu}{i \hbar}  \Big) E_j(t_1) e^{-i \epsilon (t_3-t_1)}  \Big(\frac{\alpha \hbar \omega}{i \hbar}  \Big) e^{-i(\epsilon + \omega)(t_2-t_1)} 
   \Big(\frac{\mu}{i \hbar}  \Big) E_k^* (t_2) e^{-i \omega (t-t_2)} \nonumber\\
   &\times dt_3 dt_2 dt_1  \ket{g,1}\bra{g,0}\\
 &= \alpha   \Big(\frac{\mu}{ \hbar} \Big)^2 e^{-i \omega t}  \int_{-\infty}^{t_2}  \int_{t_1}^{t} E_j(t_1) E_k^* (t_2)
 e^{-i \epsilon (t_2-t_1)} (e^{i \omega t_2} -e^{i \omega t_1}) dt_2 dt_1 \ket{g,1}\bra{g,0} \\
 &=  \alpha   \Big(\frac{\mu}{ \hbar} \Big)^2 e^{-i \omega t} e^{i \Omega (\tau_j - \tau_k)}  \int_{-\infty}^{t_2}  \int_{t_1}^{t} \exp \Big(-\frac{(t_1-\tau_j)^2+(t_2-\tau_k)^2}{\sigma^2} \Big) 
 e^{-i(\epsilon-\Omega)(t_2-t_1)} \nonumber\\
 &\times  (e^{i\omega t_2} - e^{i \omega t_1}) dt_2 dt_1 \ket{g,1}\bra{g,0}
 \end{align}

 In off-resonance, the time the system spends in an electronic coherence is limited by the Heisenberg relation ($\Delta t \approx 1/\Delta \omega$, with a detuning $\Delta \omega = (\epsilon - \Omega)$).\cite{Mukamel1995}
Then the coherent time of the polarization is short and $t_1$ is approximated by $t_1 = t_2 -\delta$ with a small value $\delta$.
Then we get
\begin{align}
e^{i\omega t_2} - e^{i\omega t_1} = e^{i\omega t_2} (1-e^{-i \omega \delta}) \approx (-i \omega \delta) e^{i \omega t_2}.
\end{align}

 The integral part is calculated as
\begin{align}
S_{jk} &=  \int_{-\infty}^{t_2}  \int_{t_1}^{t} \exp \Big(-\frac{(t_1-\tau_j)^2+(t_2-\tau_k)^2}{\sigma^2} \Big) 
 e^{-i(\epsilon-\Omega)(t_2-t_1)} (e^{i\omega t_2} - e^{i \omega t_1}) dt_2 dt_1 \\
 &\approx  (-i \omega \delta) e^{-i (\epsilon -\Omega)\delta}  \int_{-\infty}^{\infty}  \exp \Big(-\frac{(t_2 - \delta -\tau_j)^2+(t_2-\tau_k)^2}{\sigma^2} \Big)  e^{i\omega t_2} dt_2 \\
 &=  (-i \omega \delta) e^{-i (\epsilon -\Omega)\delta} \sqrt{\frac{\pi}{2}} \sigma \exp \Big(-\frac{\omega^2 \sigma^2}{8} \Big)
 \exp \Big(\frac{i \omega (\delta +\tau_j +\tau_k)}{2} \Big) \exp \Big( -\frac{(\tau_k -\tau_j -\delta)^2}{2 \sigma^2} \Big),
\end{align}
 where  the integral was determined in an approximation by expanding the upper limit to $\infty$.
 
 Then the density operator is obtained as
 \begin{align}
\hat{\rho}_{jk}(t) = & \alpha \Big( \frac{\mu}{\hbar} \Big)^2 
  \sqrt{\frac{\pi}{2}} \sigma \exp \Big(-\frac{\omega^2 \sigma^2}{8} \Big)
   (-i \omega \delta) \exp(-i (\epsilon -\Omega)\delta)
   \exp (-i \omega t) \nonumber\\
   & \times \exp (i \Omega (\tau_j - \tau_k))
  \exp \Big(\frac{i \omega (\delta +\tau_j +\tau_k)}{2} \Big) \exp \Big( -\frac{(\tau_k -\tau_j -\delta)^2}{2 \sigma^2} \Big)  \ket{g1}\bra{g0} \\
  \approx & \alpha \Big( \frac{\mu}{\hbar} \Big)^2 
  \sqrt{\frac{\pi}{2}} \sigma \exp \Big(-\frac{\omega^2 \sigma^2}{8} \Big) \nonumber\\
   & \times    (-i \omega \delta)  \exp (i \Omega (\tau_j - \tau_k))
  \exp \Big(- i \omega (t-\frac{\tau_j +\tau_k}{2}+ \theta) \Big) \exp \Big( -\frac{(\tau_k -\tau_j)^2}{2 \sigma^2} \Big) \ket{g1}\bra{g0} \\
  =&  (-i \omega \delta) A  \exp (i \Omega (\tau_j - \tau_k))
  \exp \Big(- i \omega (t-\frac{\tau_j +\tau_k}{2} + \theta) \Big) \exp \Big( -\frac{(\tau_k -\tau_j)^2}{2 \sigma^2} \Big)  \ket{g1}\bra{g0},
 \end{align}
 where the phase $\theta$ is defined as $\theta = (\epsilon -\Omega)\delta/\omega$, and
 the real value $A$ is defined as
  \begin{align}
  A = \alpha \Big( \frac{\mu}{\hbar} \Big)^2   \sqrt{\frac{\pi}{2}} \sigma \exp \Big(-\frac{\omega^2 \sigma^2}{8} \Big).
  \end{align}
 
 At $j=k=1$, $\tau_1=\tau_2 =0$ and $\hat{\rho}_{11} (t)$ is obtained
   \begin{align}
\hat{\rho}_{11} (t) = -i \omega \delta \exp(-i \omega t+ \theta) \ket{g1}\bra{g0}.
  \end{align}
  
  At $j=k=2$, $\tau_1=\tau_2 = \tau$ and $\hat{\rho}_{22} (t)$ is obtained
   \begin{align}
\hat{\rho}_{22} (t) = -i \omega \delta \exp(-i \omega (t-\tau + \theta)) \ket{g1}\bra{g0}.
  \end{align}
  
  At $j=1$ and $k=2$, $\tau_1 = 0$, $\tau_2 = \tau$ and $\hat{\rho}_{12} (t)$ is obtained
   \begin{align}
\hat{\rho}_{12} (t) = -i \omega \delta \exp(-i \omega (t- \frac{\tau}{2} + \theta)) \exp(- i \Omega \tau)
\exp (-\frac{\tau^2}{2 \sigma^2}) \ket{g1}\bra{g0}.
  \end{align}
  
  At $j=2$ and $k=1$, $\tau_1 = \tau$, $\tau_2 = 0$ and $\hat{\rho}_{21} (t)$ is obtained
   \begin{align}
\hat{\rho}_{21} (t) = -i \omega \delta \exp(-i \omega (t- \frac{\tau}{2}+ \theta)) \exp(i\Omega \tau)
\exp (-\frac{\tau^2}{2 \sigma^2}) \ket{g1}\bra{g0}.
  \end{align}

Then we get
 \begin{align}
\sum_{j,k=1}^2 \hat{\rho}_{jk} (t) =& -i \omega \delta e^{-i \omega (t+ \theta)} \Big(1+e^{i \omega \tau} + e^{i \omega \tau/2} 2 \cos (\Omega \tau) \exp (-\frac{\tau^2}{2 \sigma^2})  \Big) \ket{g1}\bra{g0} \\
=& -i \omega \delta e^{-i \omega (t+ \theta)} \Big(1+e^{i \omega \tau} + B e^{i \omega \tau/2}  \Big) \ket{g1}\bra{g0}\\
=&  -i \omega \delta \exp \Big(-i \omega (t- \frac{\tau}{2}+ \theta) \Big) \Big(2 \cos (\frac{\omega \tau}{2}) + B \Big) \ket{g1}\bra{g0}
  \end{align}
where the real value $B$ is defined as
  \begin{align}
  B = 2 \cos (\Omega \tau) \exp (-\frac{\tau^2}{2 \sigma^2}).
  \end{align}
  The Hermitian conjugate of the density operator is
  \begin{align}
\sum_{j,k=1}^2 \hat{\rho}_{jk}^\dagger (t) 
=&  i \omega \delta \exp \Big(i \omega (t- \frac{\tau}{2}+ \theta) \Big) \Big(2 \cos (\frac{\omega \tau}{2}) + B \Big) \ket{g0}\bra{g1}.
  \end{align}
  The total density operator $\rho (t)$ is
 \begin{align}
\hat{\rho} (t) = \sum_{j,k=1}^2 \Big(  \hat{\rho}_{jk} (t) + \hat{\rho}_{jk}^\dagger (t) \Big).
  \end{align}
  
 The atomic displacement is expressed by an operator $\hat{Q} = C(\hat{b}^\dagger + \hat{b})$, where $C$ is the constant of proportionality.
The expectation value of the atomic displacement is obtained by 
 \begin{align}
\braket{Q(t)} =& {\rm{Tr}} [\hat{Q} \hat{\rho} (t)] \nonumber\\
=& C \Big(2 \cos (\frac{\omega \tau}{2}) + B \Big) (i \omega \delta)
\Big( \exp \Big(i \omega (t- \frac{\tau}{2}+ \theta) \Big) - \exp \Big( -(i \omega (t- \frac{\tau}{2}+ \theta) \Big) \Big) \\
=&  2C \omega \delta  \Big(2 \cos (\frac{\omega \tau}{2}) + B \Big) \sin \Big(\omega (t - \frac{\tau}{2}+ \theta) \Big).
 \end{align}
 Then it oscillates with an angular frequency of $\omega$ and a phase of $(-\omega \tau/2+\theta)$: $\braket{Q(t)} = Q_0 \sin (\omega t -\omega \tau/2+\theta)$.
 The phonon amplitude $Q_0$ is obtained by
  \begin{align}
Q_0 =&  \Big| 2C \omega \delta  \Big(2 \cos (\frac{\omega \tau}{2}) + B \Big) \Big| = 2C\omega \delta \sqrt{2(1+\cos (\omega \tau)) + 4B \cos (\omega \tau/2) + B^2}.
 \end{align}


\end{document}